# High Performance Atomically Thin Flat Lenses

Han Lin[1#], Zai-Quan Xu[2#], Chengwei Qiu[3], Baohua Jia[1*], Qiaoliang Bao[2*]

1. *Centre for Micro-Photonics and CUDOS, Faculty of Science, Engineering and Technology, Swinburne University of Technology, P. O. Box 218, Hawthorn VIC 3122, Australia*

2. *Department of Materials Science and Engineering, Faculty of Engineering, Monash University, Clayton 3800, Victoria, Australia*

3. *Department of Electrical and Computer Engineering, National University of Singapore*

[#] These authors contribute equally to this project.
* E-mail: Qiaoliang.bao@monash.edu.au; bjia@swin.edu.au

**Abstract**

We experimentally demonstrate ultrathin flat lenses with a thickness of 7 Å, which corresponds to the fundamental physical limit of the thickness of the material, is fabricated in a large area, monolayer, CVD-prepared tungsten chalcogenides single crystals using the low-cost flexible laser writing method. The lenses apply the ultra-high refractive index to introduce abrupt amplitude modulation of the incident light to achieve three-dimensional (3D) focusing diffraction-limited resolution (0.5λ) and a focusing efficiency as high as 31%. An analytical physical model based diffraction theory is derived to simulate the focusing process, which shows excellent agreement with the experimental results.

**Introductions**

Ultrathin flat lenses that are able to focus optical energy with minimal aberration have attracted great attention as essential optical components in nano-optics and on-chip photonic systems. Recently several innovative concepts, such as metasurface[1], metamaterial[2] and super-oscillations[3], have been developed to demonstrate flat lenses with thicknesses of several tens to several hundreds of nanometers. However, those designs exploit the absorption or scattering of the incident light from metal materials or dielectric materials with high optical absorption, which results in extremely low efficiency (typically less than 1%). Further, fabricating such nanostructures relies heavily on the highly costly and time consuming tools with complex multistep processes, *i.e.*, Electron beam lithography (EBL) and focused ion beam (FIB) milling. Recently, considerable efforts have been devoted to reduce transmission loss and improve the focusing efficiency of the flat lenses by using thinner materials or optical transparent dielectric material to fulfill the stringent demands in nano-optics and on-chip photonic systems. Recent research on the two-dimensional (2D) materials, *i.e.*, graphene, has cast light on reduction of the dimension of the flat lenses.[4] It is recently demonstrated a highly efficient (focus efficiency up to 32%) ultra-thin flat lens based on graphene oxide (GO) materials with 200 nm thickness.[5] The thinnest lenses based on monolayer graphene have been demonstrated by Butt *et al.,* however, the focusing efficiency is limited to 3%.[6]

However, the insufficient phase or amplitude modulation based on the intrinsic refractive index and low absorption of the materials results in poor performance when the thickness is reduced to sub-nanometer. Therefore, it is required to introduce sufficient phase or amplitude modulation to achieve high performance ultra-thin flat lens by adapting materials with high refractive indices. Further, practical applications of such optical devices will also demand improving the efficiency and reducing the cost of production, through (1) new growth methods to prepare materials with suitable dimensions and optical properties, and (2) simple and scalable fabrication technologies.

2D atomic layered materials, *e.g.*, $MoS_2$, $WS_2$ and $WSe_2$, have been intensively studied as next-generation candidates for nanometric optoelectronic devices due to their strong light-matter interactions from the ultra-thin nature.[4,7-11] For example, bulk $WSe_2$ crystal, is built up of van der Waals bonded Se-W-Se brick units. Within each unit, two hexagonal arranged Se atom planes are coordinated with a hexagonal arranged W atom plane through covalent bond. P-type semiconducting monolayer (1L) $WSe_2$ is less than 1nm thick, and it exhibits a direct band gap of ~1.65 eV, with ~10 % absorption at its resonant peak. Also its bulk counterpart is a layered indirect semiconducting material, with a band gap of 1.2 eV.[12,13] Moreover, the unique optical properties of monolayer TMDs, namely the extraordinary large value of refractive index is as remarkably high as 5.5.[14] Recently, Lu *et al.* have pushed the thickness of an optical lens to several atomic layers using $MoS_2$ material[15], yet the efficiency is extremely low (0.3%).

In this paper, an ultrathin flat lens with a thickness of 7 Å, which corresponds to the fundamental physical limit of material thickness, is fabricated using direct femtosecond laser writing on a large single crystal $WSe_2$ monolayer prepared with chemical vapor depositions system. In contrast to the conventional phase modulation principle, the three-dimensional (3D) focusing with diffraction-limited resolution (0.5λ) and a focusing efficiency that is as high as 31% were achieved through abruptly modulate the amplitude of the incident light. More importantly, such a diffractive design is able to tune the focal length and numerical aperture of lens by varying the geometric structures instead of surface curvature. Besides, the monolayer TMD crystals can be transferred to arbitrary substrates with nondestructive polystyrene mediated transfer technique[16], making our lens readily compatible with diverse electronic or photonic devices. Our lens offers wide potential applications in on-chip nanophotonics, such as imaging, data storage and communications.

**Conceptual design of the monolayer WSe$_2$ lens.** A 3D schematic image of the monolayer WSe$_2$ single crystal based flat lens is shown in Fig. 1a, where the monolayer WSe$_2$ crystals are grown on quartz substrates with a triangular shape due to its hexagonal atomic arrangement. Inset Fig. 1a shows the atomic structure of the monolayer WSe$_2$ crystal, where a layer of W atoms are sandwiched between two layers of Se atoms. The WSe$_2$ lens is comprised of several submicrometer concentric rings, which are fabricated using the direct laser writing (DLW) method. The diffractive design relies on the interference of light from different part of the optical element, therefore sufficient amplitude or phase modulation of incident light is necessary to maximize the constructive interference and minimize distractive interference for high focusing performance. The phase modulation φ and the amplitude modulation α due to the absorption of the material can be expressed as $\varphi = nd/\lambda$ and $\alpha = 4\pi kd/\lambda$ where $n$ is the refractive index and $k$ is the extinction coefficient, respectively. For the material with sub-nanometer thickness ($d \ll \lambda$), the intrinsic phase and amplitude modulations are negligible. Thus, it is especially challenging for subnanometer-thick materials to achieve high focusing performance with a flat design.

The model of the abrupt amplitude modulation is shown in Fig. 1b, where a layer of ultrathin material with ultra-high refractive index $n_2$ is sandwiched between two media with refractive indices $n_1$ and $n_3$, respectively. The electric fields of incident light and transmitted light are $E_0 = A_0 e^{-i\varphi_0}$ and $E_1 = A_1 e^{-i\varphi_1}$, respectively, where A stands for amplitude transmission and φ stand for phase change. It has been demonstrated that the Fresnel coefficients can be derived based on the quantum mechanical principles[17], where the light is considered as waves of particles of photons. When the incident light hit on the WSe$_2$ monolayer, the photons interact with the electrons in the material, a proportion of photons that hit on the atoms are either scattered or reflected, while the others pass through the material. The probability of photons polarizing along the $p$ and $s$ direction transmitting through the interface $n_1$ and $n_2$ which can be calculated using Fresnel equations[18], are $Ts_{12}$ and $Tp_{12}$, respectively. And the probability of photons transmitting through the $n_2$ and $n_3$ interface are $Ts_{23}$ and $Tp_{23}$, respectively.

$$\frac{A_1}{A_0} = \sqrt{(Ts_{12} \cdot Ts_{23} + Tp_{12} \cdot Tp_{23})} \ .$$

The normalized amplitude transmission $A_1/A_0$ is a function of the refractive index of $n_1$, $n_2$ and $n_3$, as shown in fig. 1c. In the experiment, light propagate along the direction that is perpendicular to the substrate surface and is focused by the lens in Air. To achieve optimal interference, it is required to minimize the amplitude transmission. As for monolayer graphene ($n_2 = 2.6$), the amplitude transmission is around 90%, in contrast, the monolayer $WSe_2$ is able to get an amplitude transmission down to around 60%, which corresponding to the intensity transmission of 36%, which will greatly enhance the lens performance.

**Synthesis and characterizations of large area single crystalline $WSe_2$ monolayers.** The optical images of 1L $WSe_2$ crystals grown on a quartz and $SiO_2$/Si substrate *via* atmospheric pressure chemical vapour deposition (APCVD) are shown in Fig. 2a and b, respectively.[10] The uniform contrast from individual crystal suggests an optically flat surface and these crystals are mostly equilateral triangular with a side length of up to 19 μm on quartz and up to 69 μm on $SiO_2$/Si substrate.[19] The topography of the $WSe_2$ flakes grown on quartz substrates is investigated with the atomic force microscopy (AFM), as shown in Fig. 2c. The whole crystal is nearly atomically flat with a $SiO_2/WSe_2$ step of ~ 7Å, confirming the thickness to be monolayer. Fig. 2d plots the Raman spectra of 1L and 2L $WSe_2$. A predominating peak at 249 cm$^{-1}$, both observed on 1L and 2L $WSe_2$ crystals, is assigned to the $E^1_{2g}$ mode. The peak at 308 cm$^{-1}$, which is related to the interlayer interaction, is only observed in 2L WSe$_2$ samples. Other peaks at 358 and 374 cm$^{-1}$, attributed to the 2E$_{1g}$ and A$_{1g}$ + LA modes as reported before, are also identified.[20,21] Fig. 2e shows a Raman $E^1_{2g}$ band intensity image of a 1L WSe$_2$, the relatively uniform intensity over the whole flake suggests that the sample maintains good uniformity. The brighter spot in the centre of the crystals is assumed to be the nucleate sites.[22] A refractive index (n) as high as 5.5 (@ 633nm ) obtained from monolayer WSe$_2$, as shown in Fig. 2f, is measured using ellipsometry and extract by fitting with a broadened Lorentzian.

**Theoretical design of the monolayer WSe$_2$ lens.** To elucidate the light interaction with the WSe$_2$ flat lens, an analytical model based on the Rayleigh-Sommerfeld diffraction theory[15] is developed to evaluate the focusing capability of the WSe$_2$ lens. When a uniform plane wave impinges on the WSe$_2$ lens, the beam that is reflected by the WSe$_2$ experienced abrupt amplitude modulation while the beam propagating through the ablated zones only experiences ignorable amplitude modulations. As such, the interferences effects from wavelets in the lens plane merged the transmitted light into a 3D focusing "corn"[18], as illustrated in Fig. 3a. The refractive index distribution in the lens plane is shown in Fig. 3b, which shows a concentric ring patterns of air grooves in the high index WSe$_2$ crystal. The corresponding amplitude and phase modulation generated when the lights go through the WSe$_2$ lens are shown in Fig. 1c. Because of the ultra-thin nature of the WSe$_2$ monolayer, the maximum phase change between adjacent air and WSe$_2$ zones is extremely small (~0.04π), even though the refractive index difference between WSe$_2$ and air is giant (Δ$n$ is 4.5 at the wavelength of 633 nm). On the other hand, the difference in the amplitude transmission between the air and WSe$_2$ zones is substantial (up to 40%).

To consider both the amplitude and phase modulation enabling the optimal focusing condition, the radius of the m$^{th}$ ablated zone $a_m$ need to satisfactory the following condition[23]

$$a_m = \sqrt{\lambda f (2m - \Delta\varphi/\pi)}, \tag{1}$$

where $f$ is the designed focal length of the lens (Fig. 3a) and $\Delta\varphi$ is the phase difference between the air and WSe$_2$ zones, $\lambda$ is the wavelength of the incident light. In the case of monolayer WSe$_2$, the $\Delta\varphi \approx 0$, thus it is possible to ignore the phase modulation in the lens design. Therefore, the equation (1) is simplified to

$$a_m = \sqrt{2m\lambda f} \tag{2}$$

It is thus evident that $a_1 = \sqrt{2\lambda f}$ determines the positions of the rest air zones and the focal length of the lens. As a result, the focusing principle of the lens is the abrupt amplitude modulation result from the high refractive index of the 2D materials.

Additionally, the effective numerical aperture (NA) of the lens can be modulated by the number of rings ($N$) since the overall radius of the lens is defined by the radius of the outmost ring $a_N = \sqrt{2N\lambda f}$. Therefore, the effective NA of the lens can be calculated as

$$NA = \frac{a_N}{f} = \sqrt{\frac{2N\lambda}{f}} \tag{3}$$

Therefore, it is possible to achieve maximal $NA = 1$, by using $N=ceil(f/2\lambda)$, where $ceil(x)$ is the function to round the element $x$ to the nearest integer towards infinity. In experiments, the $N$ is limited by the area of the monolayer WSe$_2$ crystal experimentally though. Another factor affecting the geometry of the lens is the line width of each ring ($l$) (Fig. 3b). The line width which separates two adjacent rings, is required to be smaller than the distance between the two rings as $l<a_{m+1}-a_m$. On the other hand, as $\sin\beta \propto 1/(a_N - a_{N-1})$, the distance between the two outmost rings determines the maximum convergence angle $\beta$ (Fig. 3a) of the lens as well. Therefore, the minimal line width of each air zone also decides the largest $N$, which governs the NA of the lens. As such, $a_1$, $l$ and $N$ together are used to define the geometry of the monolayer WSe$_2$ lenses, and the effects of these parameters on lens performance are discussed later. The cross-sectional plots along the x and z directions calculated with $a_1 = 3.5$ μm, $N = 5$, $l = 400$ nm, are shown in Fig. 3d and 3e, respectively. And the calculated intensity distributions of the focal spot is shown in Fig.3f, which is very strong and well-defined with ignorable side-lobes (< 10% of the intensity of the central lobe). The full width at a half maximum (FWHM) of the focal spot in the x-axis direction ($W_x$) is ~ 0.51λ and that in the z-axis direction ($W_z$) is ~1.54λ, respectively, forming a high-quality 3D focal spot with wavelength-scale resolution.

**Fabrication of the monolayer $WSe_2$ lenses**. To verify the simulation results, lenses are fabricated *via* femtosecond laser enabled two-photon ablation of the $WSe_2$ material, as schematically shown in Fig. 4a. The $WSe_2$ sample is mounted on a 3D nanometric scanning stage (Physik instrument), which allows scanning the laser focal spot relative to the sample, as such, arbitrary pattern with nanometer positioning accuracy can be fabricated. The ring width (*l*) is controlled by the dimension and power of the laser spot. As mentioned before small *l* (down to submicrometer scale) is key to achieve high focusing efficiency. However, the minimum line width achieved with continuous wave laser is larger than 1 μm due to high thermal conductivity of the TMD materials in the 2D plane[24], which is not suitable for achieving subwavelength resolution. Therefore, to minimize the linear photo-thermal effect and absorption, a low repetition rate near infrared femtosecond laser (100 fs, 10 KHz, 800 nm) is chosen (Fig. 4b). As a result, a record small feature size of ~ 300 nm (0.38λ) is achieved through high NA (0.95) objective, which represents a more than three-fold resolution enhancement compared to the literature.[24] The reflection optical microscopic image of a fabricated $WSe_2$ lens on $SiO_2$ substrate is shown in Fig. 4c, in which $WSe_2$ (white part) and milled $WSe_2$ (black part) are clearly resolved due to the high reflection contrast. The phase modulation is measured by using the optical profiler to measure the optical path length (OPL) difference between the air and $WSe_2$ zones. The measured OPL step between the $WSe_2$ and the ablated parts is ~ 10 nm, corresponding to a change of ~5 in refractive index (Supplementary material). The complete removal of the WSe$_2$ material in the patterned area is further confirmed by a Raman spectra and E$^1_{2g}$ band intensity imaging (Fig. 4d). The shrink of the E$^1_{2g}$ and the rising peak at 308cm$^{-1}$ suggest that WSe$_2$ is converted to WO$_x$ after laser milling (Supplementary material). AFM was used to study morphology of the lens. As shown in Fig 4e, the particles with a diameter of ~ 73 nm were observed on the milled part, meaning the resultant WO$_x$ is amorphous and isolated. The root-mean-square (RMS) of milled area is ~ 0.59 nm, compared with ~ 0.26 nm for as-grown part. The cross-sectional phase profile is shown in the inset, the 10° phase difference can be identified in the lower panel of the figure and the height step measured is around 0.6 nm. Besides, the line width can be accurately controlled by the laser power. As shown in Fig. 4f, the line width from 300 nm up to 950 nm can be continuously tuned by varying the laser power from 3 µW to 14 µW. In addition, the data shows an exponential trend, which confirms the ablation is based on nonlinear effect.

**Characterization of the monolayer WSe$_2$ lens**. To characterize the performance of the monolayer WSe$_2$ lenses, the fabricated lenses are mounted on a 1D nanometric scanning stage along the axial direction. The lenses are illuminated by a collimated laser beam at 633 nm. The intensity distribution in the focal region is imaged and magnified using a 4$f$ image system composed of an high NA objective (100×, 0.95NA) and a tube lens (200 nm) on to the CCD camera. Each lens is characterized by recording a serial of CCD images along the z-direction with a 10 nm step. In this way, the focal plane can be accurately identified by the position at the peak focal intensity. The cross-sectional images of the focal spot in the x-y plane and x-z plane are shown in Fig.5 b and c, respectively. The cross-sectional plots (marked using dash line in Fig. 5b and c) are shown in the Fig. 5d and e, respectively, which are comparable with the theoretical prediction. The W$_x$ and W$_z$ are slightly larger than the theoretical results due to scattering of the lights from the particles on the substrate. The experimental 3D focal spot is reconstructed by stacking and merging all the CCD images captured, as presented in Fig. 5f. The surface of the 3D focal spot is smooth and $W_x$ and $W_z$ are $0.58\lambda$ and $1.75\lambda$, respectively, which agrees well with the simulated results.

To investigate the relationship between the focal length of the $WSe_2$ lenses and $a_1$, lenses with $a_1$ ranging from 1.37 μm to 3.89 μm, $N = 5$ and $l = 400$ nm were fabricated and tested. The resulted dependence of the focal length on $a_1$ is shown in Fig. 5g, which can be fitted with a parabolic function as predicted using Eq. (2). When $a_1$ is less 2 μm, peak intensity of the focal spot is too low to be distinguished from surroundings due to small light collecting area from the lens. Further, the $W_x$ and $W_z$ as a function of $a_1$ are plotted in Fig. 5h. The FWHMs are almost unchanged against $a_1$, since the effective NA of the lenses depends only on the $N$ as predicted by Eq. (3).

Finally, lenses with different N (N varies from 3 to 8 with $a_1$ is fixed at 3.89 μm) were fabricated to study the relationship between NA and N. It is found the lens requires at least 3 rings to achieve decent focusing, as shown in Fig. 6a. The relationship between N and $W_x$, $W_z$, is depicted in Fig. 6b, the diffraction-limit performance of the lateral resolution (around $0.5\lambda$) can be achieved when N is larger than 5. On the other hand, the axial resolution can be continuously improved by increasing the number of rings as the light is strongly diffracted by the outer rings which results in a large convergence angle β. In addition, higher N results from a larger overall size of the lenses allows guiding more lights to the focal spot. Consequently, the peak intensity of the focal spot is increasing along with the increasing of the N (Fig. 6c). The focal length of the lens that depends on $a_1$ remains unchanged (Fig. 6d). Therefore, it is possible to tuning the focal length and the effective NA of the lens separately by controlling the $a_1$ and $N$, respectively. Table I summarized and compared the existing lens based on 2D materials, the performance that our lens exhibit is the highest among those with sub-nanometer thick lenses.

**Conclusions**

We have demonstrated the new ultrathin flat lens based on the abrupt amplitude modulation enabled by the high refractive index of the monolayer $WSe_2$ crystal, which is able to achieve diffraction-limit focusing performance in far field. A theoretical model is developed to understand and predict the focusing performance of the lenses. The lenses are fabricated *via* two-photon femtosecond laser ablation of the $WSe_2$ material. The focusing properties of the lenses, *e.g.,* focal length and NA, can be tuned by control the radius and number of the rings. Our lenses are ultralight weight, highly efficient and compatible, which open up new revenues for various multidisciplinary applications including non-invasive 3D biomedical imaging, laser tweezing, all-optical broadband photonic chips, light harvesting, aerospace photonics, optical microelectromechanical systems and lab-on-chip devices.

# Experimental

**Synthesis and Characterizations of monolayer WSe$_2$ single crystals.**

Monolayer WSe$_2$ crystals are prepared using the atmospheric pressure chemical vapor deposition (APCVD) method on quartz, SiO$_2$/Si and c-sapphire (0001) substrates, as reported previously.[10] Raman and PL measurements were performed using confocal microscope system (WITec, alpha 300R) with a 100× objective under ambient conditions. The signals were collected by the spectrometer equipped with a 1800 line/mm grating. The samples were placed on a piezo crystal-controlled scanning stage and excited with a 532 nm laser. Low laser power (50 µW) was applied during the measurements to avoid damaging the monolayers. The topography WSe$_2$ lenses were characterized using AFM (Bruker, Dimension Icon SPM) in tapping mode. The optical path length characterizations are obtained with an interferometer (Bruker Contour GT-I) in a phase-shifting mode. An ellipsometer (J.A. Woollam M-2000) has been used to measure optical parameters of the monolayer WSe$_2$ material.

**Theoretical design and numerical modelling.**

The analytical model is developed based on the Rayleigh-Sommerfeld diffraction theory. The focal intensity distribution of the lenses is calculated using a homemade Matlab program.

**Femtosecond laser fabrication and imaging characterization.**

A DLW system is used to fabricate the monolayer WSe$_2$ lens, as reported previously[27]. A low repetition rate femtosecond pulsed laser beam (100 fs pulse, 10 KHz, 800 nm) is used to minimize the thermal effect. A homemade imaging characterization system is built to study the performance of lenses. The cross-sectional distributions of the generated focal spots of the WSe$_2$ lenses are captured using a CCD camera equipped with a 100× objective (NA = 0.95). The magnification rate of the 4f image system is 110. Therefore, each CCD pixel corresponds to a step size of 63 nm in the lateral direction. The 3D images of the focal spots are reconstructed by staking images captured at different len-to-objective distances.

**Author's contributions**

L. H. contributes to the lens fabrications, performance characterizations and simulations; Z. X. contributes to the materials preparations, characterizations and lens fabrications; B. J. and Q. B. conceived the idea and supervised the project. C. Q contributed to the simulations. All the authors discuss and write the manuscript.

**Acknowledgement**

Baohua Jia acknowledges support from the Australian Research Council through the Discovery Early Career Researcher Award scheme (DE120100291). This work was supported by ARC DECRA (DE120101569) and DSI top-up grant, DP (DP140101501), Engineering Seed Fund in Monash University. Z. Xu acknowledges supports from the Australian Postgraduate Award (APA) and international postgraduate research scholarship (IPRS). Dr. Q. Bao also acknowledges the support from 863 Program (Grant No. 2013AA031903), the youth 973 program (2015CB932700), the National Natural Science Foundation of China (Grant No. 51222208, 51290273, 91433107). This work was performed in part at the Melbourne Centre for Nanofabrication (MCN) in the Victorian Node of the Australian National Fabrication Facility (ANFF).

**Figure captions**

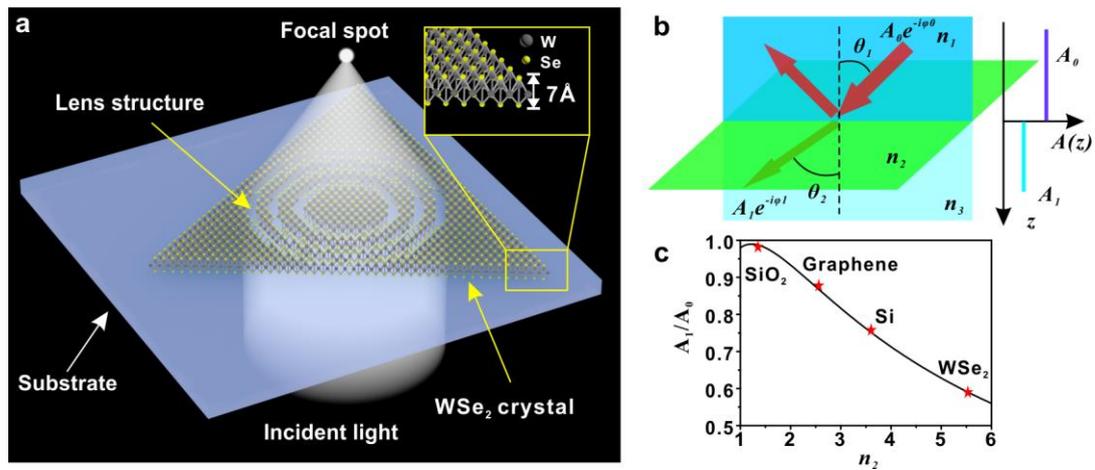

Figure. 1 **Conceptual design of monolayer WSe$_2$ lens** (a) A 3D schematic view of light focusing with monolayer WSe$_2$ lens. Inset: The atomic structure of the monolayer WSe$_2$. (b) Light transmission through an ultrathin 2D material/air interface. (c) Amplitude transmission as a function of the refractive index (n$_2$) of the 2D material.

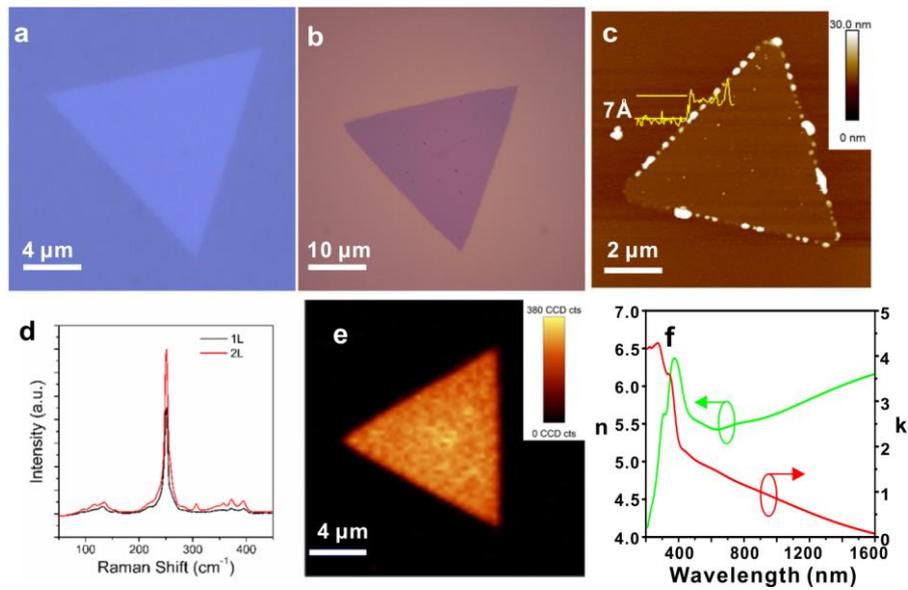

Figure 2. **Synthesis and characterization of monolayer WSe$_2$ crystal.** Optical microscopy images of a WSe$_2$ single crystal grown on quartz substrate (a) and SiO$_2$/Si substrate (b). Scale bar: 4 μm in (a) and 10 μm in (b). (c) AFM height image of a WSe$_2$ crystal grown on quartz. Scale bar: 2 μm. Inset: height profile of this WSe$_2$ crystal. (d) Raman spectra of the 1L and 2L WSe$_2$ crystals grown on quartz substrates. (e) Raman $E^1_{2g}$ band intensity mapping of a WSe$_2$ monolayer crystal. Scale bar: 4 μm. (f) Refractive index n and extinction coefficient k of monolayer WSe$_2$ crystal.

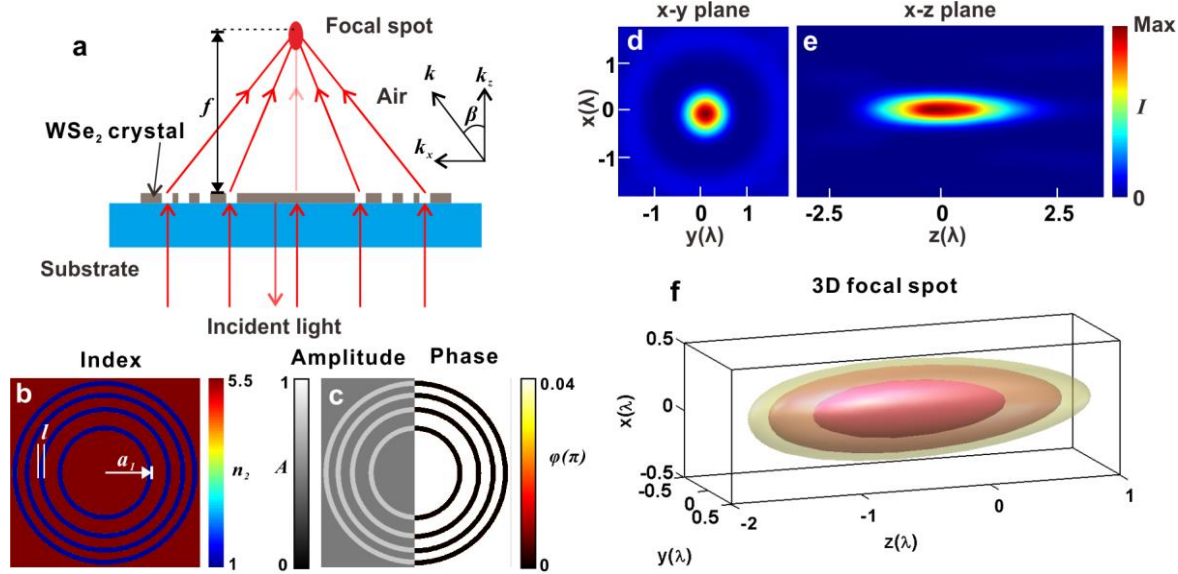

Figure 3. **Theoretical design of the monolayer WSe$_2$ lens.** (a) Schematic view of the focusing process using the monolayer WSe$_2$ lens; $f$: focal length. (b) Refractive index distribution in the lens plane, $a_1$: the radius of the inner first ring, $l$: linewidth of each ring. (c) Amplitude and phase distribution of the incident light modulated by the monolayer WSe$_2$ lens. (d) Intensity distribution of the focal spot in the x-y plane; (e) Intensity distribution of the focal spot in the x-z plane. (f) 3D surfaces of the focal spot at different intensity levels, the three shells corresponds to the intensity threshold at 0.8, 0.6 and 0.5 of the peak intensity, respectively.

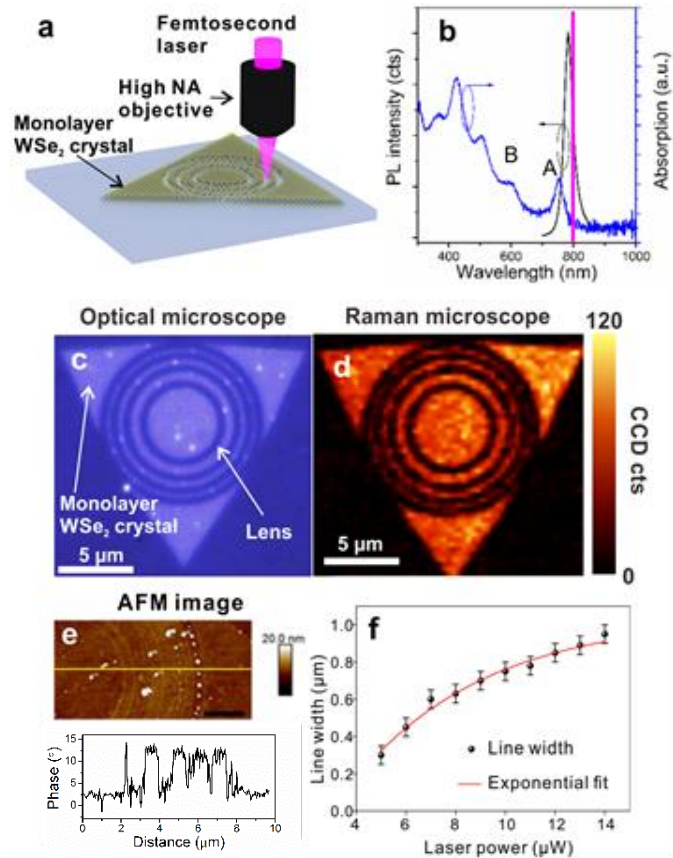

Figure 4. **Characterization of the monolayer WSe$_2$ lens fabricated by femtosecond laser fabrication.** (a) Schematic view of the femtosecond laser fabrication process of monolayer WSe$_2$ lens. (b) Absorption and photoluminescence spectra of the monolayer WSe$_2$ crystal, the wavelength of the femtosecond laser used is marked using the solid vertical line. (c) Reflective optical microscopic image of a fabricated monolayer WSe$_2$ lens. (d) Confocal Raman microscopic intensity imaging of a fabricated monolayer WSe$_2$ lens. (e) AFM image of a fabricated monolayer WSe$_2$ lens with the phase profile shown lower panel. Scale bar: 800nm (f) Line width as a function of the incident laser power.

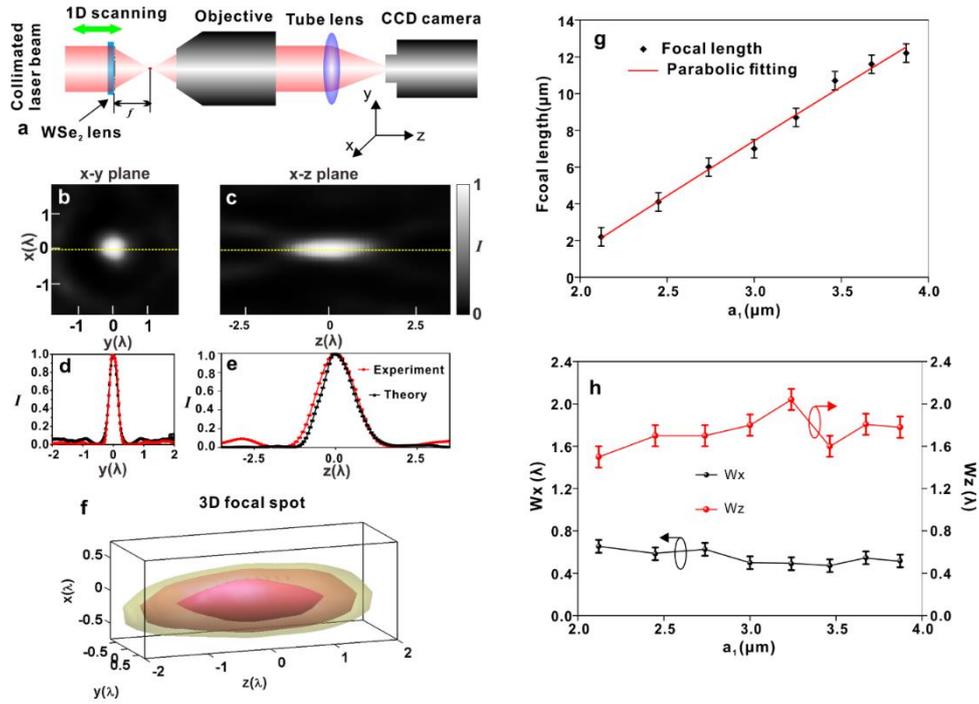

Figure 5. **Characterization of the monolayer WSe$_2$ lens.** (a) Schematic view of the characterization set up. Measured focal intensity distribution in the x-y plane (the focal plane) (b) and the x-z plane (c). Intensity cross-section plots along the y-direction (d) and z-direction (e). (f) Reconstructed 3D focal spot, the shells from inside to outsides corresponds to the 0.8, 0.6, and 0.5 of the peak intensity. (g) Focal length as a function of the radius of the first inner ring ($a_1$). (h) Curves of Wx and Wz dependence on the radius of the first ring ($a_1$).

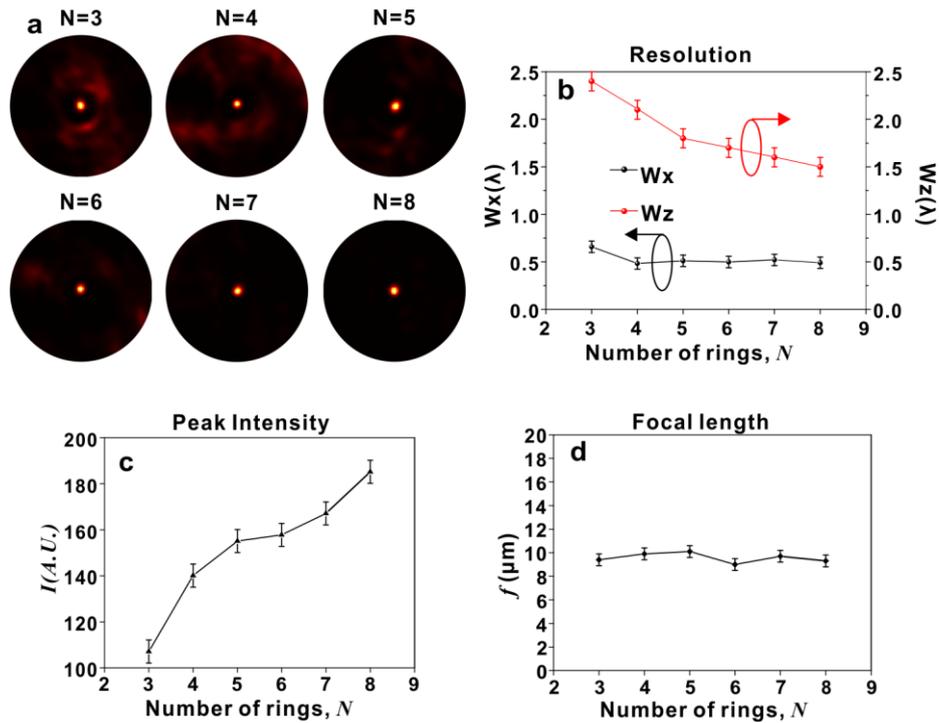

Figure 6. **The performance of the monolayer WSe$_2$ lens as a function the number of rings.** (a) Curves of FWHM along the x-direction (Wx) and the z-direction (Wz) dependence on the number of rings (*N*). (b) Curves of peak intensity dependence on the number of rings (*N*). (c) Curves of focal length dependence on the number of rings (*N*).

**Table I**, Summary and comparison of the existing lens based on 2D materials.

| Materials | Wavelength (nm) | Thickness (nm) | Focal length (μm) | FWHM ($W_x$) (μm) | FWHM ($W_z$) (μm) | Efficiency (%) |
|---|---|---|---|---|---|---|
| Graphene [6] | 800 | 1.8 | 175 | | | 2.9 |
| Graphene oxide [5] | 700 | 200 | | 0.49$\lambda$ | 0.83$\lambda$ | 32 |
| $MoS_2$ [15] | 532 | 0.7 | NA | NA | NA | 0.3 |
| | 532 | 1.4 | NA | NA | NA | 0.8 |
| | 532 | 4.2 | NA | NA | NA | 4.4 |
| | 532 | 5.6 | NA | NA | NA | 10.1 |
| $WSe_2$ (This work) | 633 | 0.7 | | 0.58$\lambda$ | 1.75 $\lambda$ | 31 |
| Gold [25] | | 400 | 6.23 | 0.88 | NA | NA |

Supporting information

## Atomically thin flat lenses

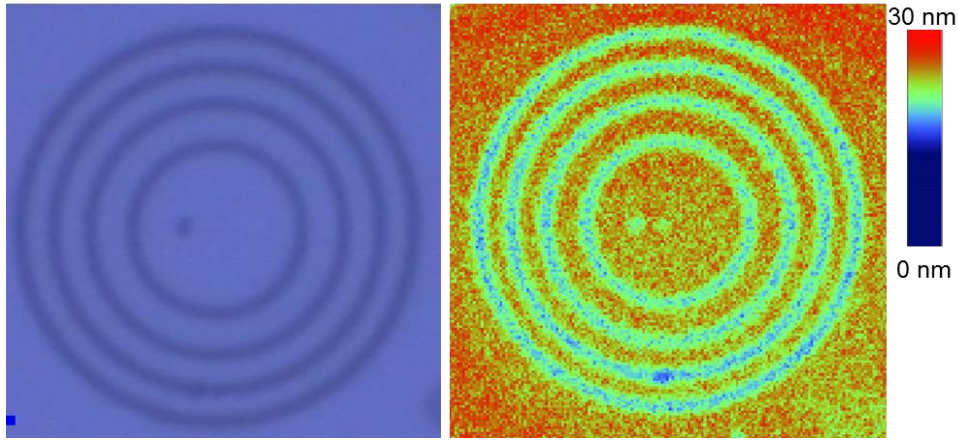

Figure S1. (a) The reflection optical microscopic image of a fabricated WSe$_2$ lens on quartz substrate. Regions with different colors correspond to WSe$_2$ (white part) and milled WSe2 (black part) are clearly resolved due to the high reflection contrast. (b) Optical path length (OPL) imaging of the WSe$_2$ lens.

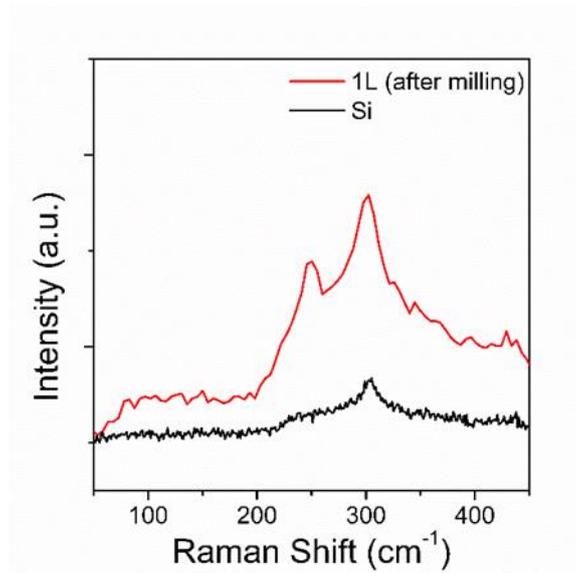

Figure S2. Raman spectra of Si and WSe$_2$ after laser milling.